\documentclass[conference]{IEEEtran}
\IEEEoverridecommandlockouts
\usepackage{cite}
\usepackage{amsmath,amssymb,amsfonts}
\usepackage{algorithmic}
\usepackage{graphicx}
\usepackage{lmodern}
\usepackage[binary-units]{siunitx}
\usepackage[abbreviations]{foreign}  % For \ie, \eg \etc
\usepackage{xcolor}
\def\BibTeX{{\rm B\kern-.05em{\sc i\kern-.025em b}\kern-.08em
    T\kern-.1667em\lower.7ex\hbox{E}\kern-.125emX}}

\usepackage{tikzexternal}
\tikzsetexternalprefix{temp/}

\title{NVMM cache design: Logging vs. Paging\\
}

\author{\IEEEauthorblockN{Rémi Dulong\IEEEauthorrefmark{1}\IEEEauthorrefmark{3}, Quentin Acher\IEEEauthorrefmark{2}, Baptiste Lepers\IEEEauthorrefmark{1}, Valerio Schiavoni\IEEEauthorrefmark{1}, Pascal Felber\IEEEauthorrefmark{1}, Gaël Thomas\IEEEauthorrefmark{3}}

\IEEEauthorblockA{\IEEEauthorrefmark{1}University of Neuchâtel, Switzerland --- \IEEEauthorrefmark{2}ENS Rennes, France --- \IEEEauthorrefmark{3}Télécom SudParis, France}
}

\begin{document}

\maketitle

\begin{abstract}
Modern NVMM is closing the gap between DRAM and persistent storage, both in terms of performance and features. 
%As every component in the memory hierarchy, caching slower devices is a use case that has to be evaluated. 
Having both byte addressability and persistence on the same device gives NVMM an unprecedented set of features, leading to the following question: How should we design an NVMM-based caching system to fully exploit its potential?
We build two caching mechanisms, NVPages and NVLog, based on two radically different design approaches. 
NVPages stores memory pages in NVMM, similar to the Linux page cache (LPC). 
NVLog uses NVMM to store a log of pending write operations to be submitted to the LPC, while it ensures reads with a small DRAM cache.
Our study shows and quantifies advantages and flaws for both designs. 
\end{abstract}

\section{Introduction}

The emergence of modern NVMM is a great opportunity to implement known designs and adapt them, or invent new ones.
We tried these two approaches with caching mechanisms for a file system stored in secondary storage. 
Indeed, caching data for slower tier storage devices (SSD or HDD) is a great use case for NVMM. 
It provides high persistence guarantees, higher read and write bandwidth and lower latencies than most persistent block devices \cite{izraelevitz:19:optane}. 
In this study, we target applications that require a high level of data consistency, which would highly solicit a regular disk with frequent calls to \texttt{fsync}. 
For such applications, we propose a persistent cache able to give fast persistence guarantees without having to wait for a slow secondary storage. 

\section{NVMM-based Caching}

\begin{figure}
  \centering
  \includegraphics[height=2.9cm]{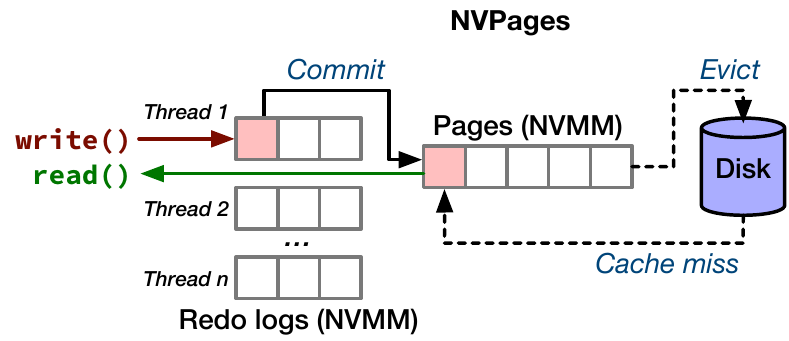}
  \vspace{-2mm}
  \caption{Core design of NVPages}
  \vspace{-2mm}
  \label{fig:nvpages}
\end{figure}

\begin{figure}
  \centering
  \includegraphics[height=4.1cm]{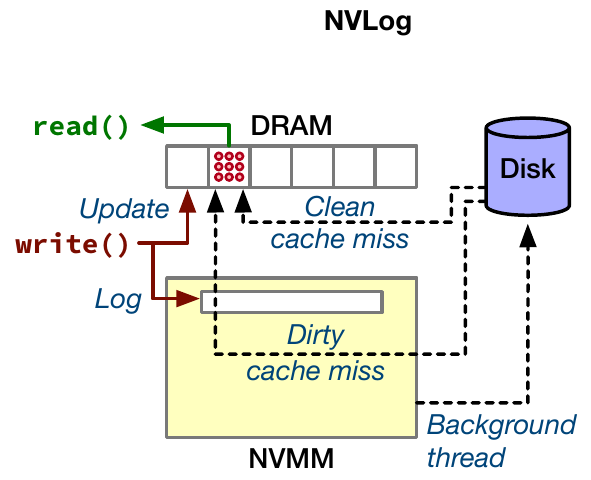}
  \vspace{-2mm}
  \caption{Core design of NVLog}
  \vspace{-2mm}
  \label{fig:nvlog}
\end{figure}

NVPages and NVLog are POSIX-like IO shared libraries. They provide standard IO functions, such as \texttt{open}, \texttt{pread}, \texttt{pwrite}, \texttt{close}, \etc.
When the shared library is loaded, NVMM is mapped, and data structures are initialized.
A flag in NVMM is set to 1 when the program is loaded, and set to 0 when it is unloaded properly.
This flag allows both caches to start a recovery procedure in case of a previous crash, flushing to disk every modification still pending in cache when the crash occurred. So far, they do not support multiple threads.
However, they differ in their core implementation, depicted in Fig.~\ref{fig:nvpages} and Fig.~\ref{fig:nvlog}.

\smallskip\textbf{NVPages.}
NVPages is designed as a regular page cache, with a few adaptations to make it compliant with NVMM and its persistence guarantees.
\SI{4}{\kibi\byte} pages are stored in NVMM.
When a page is accessed, a radix tree in volatile memory looks for a volatile metadata structure that contains a pointer to the non-volatile page.
In order to ensure consistency after a crash, calls to \texttt{pwrite} first write data in a redo log stored in persistent memory.
Then, the redo log content is flushed to the actual non-volatile page cache.
The page cache eviction is done with a least recently used (LRU) policy.
NVPages can be used in \texttt{O\_DIRECT} mode, bypassing the LPC to interact directly with the disk with aligned \SI{4}{\kibi\byte} blocks.
We do not report performance with this mode since we measured that bypassing the LPC reduces performance in read.
As described in Fig.\ref{fig:nvpages}, NVPages is designed to be adapted for multithreaded workloads. 

\smallskip\textbf{NVLog.}
NVLog builds atop NVCache \cite{dulong2021nvcache}, ported to work as a shared library. 
It embeds two main components: a NVMM log, and a small DRAM page cache.
When the \texttt{pwrite} function is called, data is written to the NVMM log. 
A background thread continuously waits for log entries and writes them to disk as soon as possible. 
To ensure consistency in this configuration, every call to \texttt{pread} should get the page from disk and check if patches (log entries) have to be applied before returning the data. 
As this would make reads very slow, NVLog keeps a small DRAM page cache (\SI{2}{\gibi\byte}) with up-to-date data. 
It also keeps track of pages that would need to be patched before returning, so it only searches in the NVMM log when necessary.
For reads, NVLog uses the LPC as an extension of NVLog's DRAM cache, from which it can fetch data instead of waiting for the disk.
For writes, NVLog submits changes to the LPC in batches, before calling fsync to ensure the data is persisted on disk.
This way, it benefits from LPC optimizations, such as merging consecutive writes on the same offset before writing the page on disk.
Its design is complex because of the internal synchronization between the application and the background thread.
Adapting it for multithread remains challenging.

%\vspace{5mm}
\smallskip\textbf{Discussion.}
NVLog is designed to absorb bursts of writes in its log, but may not be suited for mixed or parallel IOs.
It only keeps a small amount of pages updated in DRAM.
Increasing the amount of NVMM in NVLog does not change the probability of cache hit. 
Instead, NVPages is designed to maximize the probability of cache hit by keeping a lot of memory pages available in NVMM, as almost all of its allocated NVMM is dedicated to pages. 
We expected the latter approach to be more efficient for mixed IOs, reducing the amount of interactions with the disk.

\section{Evaluation}

Our benchmark machine is a Supermicro mono-socket machine with an Intel Xeon Gold 6326 CPU, 2 modules of \SI{128}{\gibi\byte} of Intel Optane v200 DCPMM \cite{intel3dxpoint}, and a \SI{512}{\giga\byte} NVMe SSD, running Ubuntu 20.04 LTS.

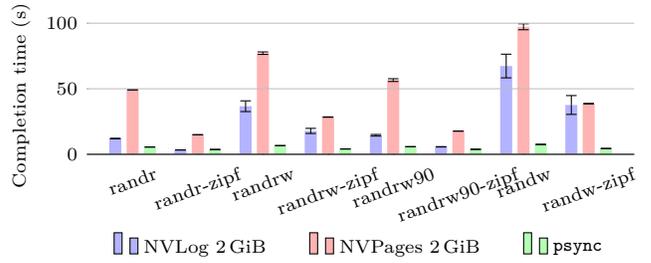
\begin{figure}
  \tikzsetnextfilename{fio2}
  \raggedleft
  \begin{tikzpicture}
  \begin{axis}[
      ybar, axis on top,
      height=3.5cm, width=\linewidth,
      bar width=0.16cm,
      ymajorgrids, tick align=inside,
      enlarge y limits={value=.1,upper},
      ymin=0, ymax=100,
      axis x line*=bottom,
      axis y line*=left,
      y axis line style={opacity=0},
      y label style={font=\scriptsize, yshift=-10, xshift=-5},
      y tick label style={font=\scriptsize},
      x tick label style={font=\scriptsize,rotate=20},
      tickwidth=0pt,
      enlarge x limits=true,
      legend style={
        draw=none,
        at={(0.5,-0.5)},
        font=\scriptsize,
        anchor=north,
        legend columns=-1,
        /tikz/every even column/.append style={column sep=0.5cm}
      },
      ylabel={Completion time (s)},
      symbolic x coords={
        randr, randr-zipf, 
        randrw, randrw-zipf,
        randrw90, randrw90-zipf,
        randw, randw-zipf,
      },
      xtick=data,
    ]
    \addplot [
      draw=none,
      fill=blue!30,
      error bars/.cd,
      y dir=both, y explicit,
    ] table[col sep=comma,header=true,x index=0,y index=1, y error index=2] {../data/fio-20G/completion-2G/nvcache.csv};

    \addplot [
      draw=none,
      fill=red!30,
      error bars/.cd,
      y dir=both, y explicit,
    ] table[col sep=comma,header=true,x index=0,y index=1, y error index=2] {../data/fio-20G/completion-2G/nvpc.csv};

    \addplot [
      draw=none,
      fill=green!30,
      error bars/.cd,
      y dir=both, y explicit,
    ] table[col sep=comma,header=true,x index=0,y index=1, y error index=2] {../data/fio-20G/completion-2G/psync.csv};

    \legend{NVLog \SI{2}{\gibi\byte}, NVPages \SI{2}{\gibi\byte}, \texttt{psync}}
  \end{axis}
\end{tikzpicture}
  \caption{FIO benchmarks with \SI{2}{\gibi\byte} of NVMM cache}
  \label{fig:FIO-2G}
\end{figure}

\begin{figure}
  \tikzsetnextfilename{fio100}
  \raggedleft
  \begin{tikzpicture}
  \begin{axis}[
      ybar, axis on top,
      height=3.5cm, width=\linewidth,
      bar width=0.16cm,
      ymajorgrids, tick align=inside,
      enlarge y limits={value=.1,upper},
      ymin=0, ymax=100,
      axis x line*=bottom,
      axis y line*=left,
      y axis line style={opacity=0},
      y label style={font=\scriptsize, yshift=-10, xshift=-5},
      y tick label style={font=\scriptsize},
      x tick label style={font=\scriptsize,rotate=20},
      tickwidth=0pt,
      enlarge x limits=true,
      legend style={
        draw=none,
        at={(0.5,-0.5)},
        font=\scriptsize,
        anchor=north,
        legend columns=-1,
        /tikz/every even column/.append style={column sep=0.5cm}
      },
      ylabel={Completion time (s)},
      symbolic x coords={
        randr, randr-zipf, 
        randrw, randrw-zipf,
        randrw90, randrw90-zipf,
        randw, randw-zipf,
      },
      xtick=data,
    ]
    \addplot [
      draw=none,
      fill=blue!30,
      error bars/.cd,
      y dir=both, y explicit,
    ] table[col sep=comma,header=true,x index=0,y index=1, y error index=2] {../data/fio-20G/completion-100G/nvcache.csv};

    \addplot [
      draw=none,
      fill=red!30,
      error bars/.cd,
      y dir=both, y explicit,
    ] table[col sep=comma,header=true,x index=0,y index=1, y error index=2] {../data/fio-20G/completion-100G/nvpc.csv};

    \addplot [
      draw=none,
      fill=green!30,
      error bars/.cd,
      y dir=both, y explicit,
    ] table[col sep=comma,header=true,x index=0,y index=1, y error index=2] {../data/fio-20G/completion-100G/psync.csv};
    
    \legend{NVLog \SI{100}{\gibi\byte}, NVPages \SI{100}{\gibi\byte}, \texttt{psync}}
  \end{axis}
\end{tikzpicture}
  \caption{FIO benchmarks with \SI{100}{\gibi\byte} of NVMM cache}
  \label{fig:FIO-100G}
\end{figure}
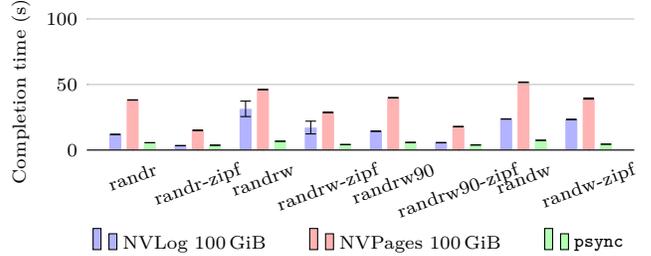

We evaluated our 2 systems with FIO \cite{axboe2005fio}.
These tests are performing \SI{20}{\gibi\byte} of random accesses through a \SI{20}{\gibi\byte}-wide file.
In Fig.~\ref{fig:FIO-2G} and Fig.~\ref{fig:FIO-100G}, we submit pure reads (\texttt{randr}), 50\% reads and 50\% writes (\texttt{randrw}), 90\% reads and 10\% writes (\texttt{randrw90}), and pure writes (\texttt{randw}).
Then, to show the efficiency of the caching policies, we measure the same tests with a Zipfian distribution\cite{zipf} that ensures 95\% of random offsets will be in 5\% of the file.
Each bar is the average completion time of 5 runs.
For each plot, we compare NVPages and NVLog with a given amount of NVMM allocated.
Our reference is the regular \texttt{psync} IO engine of FIO which uses regular POSIX functions, measuring the performance of the LPC in DRAM.
With this baseline, there is no guarantee of persistence, while NVPages and NVLog both guarantee persistence as soon as a \texttt{pwrite} call returns.
Having similar persistence guarantees with \texttt{psync} is possible, by enabling a \texttt{fsync} call after each pwrite.
However, completion times were so long that we did not include them in these plots (more than an hour for \SI{20}{\gibi\byte} of pure writes).

We expected NVPages to be less efficient in pure write workloads, because the use of redo logs leads to write every data to NVMM twice.
On the other hand, we also expected it to be more efficient than NVLog on mixed IO workloads, because it can store much more data in its page cache, increasing the cache hit probability and reducing interactions with the SSD to the minimum.

However, these results show NVLog performs significantly better in almost every workload.
The pure read performance of NVPages reveals a fundamental flaw that prevents it to perform better with mixed IOs.
By design, cache misses have a cost in NVPages, because they imply to copy the missing page to NVMM.
But the main flaw in this design is the bandwidth limitation of current NVMM compared to DRAM.
NVPages can take pages from the LPC in DRAM, but will then require to read in NVMM to retrieve them for reads or writes.
On the contrary, NVLog keeps fresh pages in DRAM, which allows us to get the full potential of DRAM read bandwidth, as we measured in Fig.~\ref{fig:FIO-2G} and Fig.~\ref{fig:FIO-100G} with \texttt{randr} and \texttt{randr-zipf} benchmarks.

\vspace{2mm}

\section{Conclusion and Future Work}

In its current state, NVLog seems to have a clear edge over NVPages.
It performed better on all workloads, even on those we expected NVPages to be more efficient.
That said, some additional logic should be added to both caches implementations in order to evaluate their performance on multithread workloads.
From a design point of view, NVPages has several advantages and may outperform NVLog on parallel IOs thanks to its independent redo logs (while NVLog must share the same log with all threads).
Furthermore, the main bottleneck we found in NVPages relies on the difference of performance between DRAM and NVMM.

\vspace{2mm}

\smallskip\textbf{Acknowledgments.} This work received funds from the Swiss National Science Foundation (FNS) under project PersiST (no. 178822).

\vspace{2mm}

\bibliographystyle{IEEEtran}
\bibliography{references}

\end{document}